\documentclass[preprint,12pt]{elsarticle}
\usepackage{amssymb}
\usepackage{amsmath}
\journal{Physics Letters A}
\newcommand{\absol}[1]{\vert{#1}\vert}
\begin{document}
\begin{frontmatter}
\title{On the  anomalous density of a dilute homogeneous Bose gas}		
\author[label1,label2]{Abdulla Rakhimov} 
\author[label3]{Mukhtorali Nishonov}
\affiliation[label1]{organization={Laboratory of Theoretical Physics, Institute of Nuclear Physics}, city={Tashkent}, postcode={100214}, country={Uzbekistan}}
\affiliation[label2]{organization={Center for Theoretical Physics, Khazar University}, city={Baku}, postcode={AZ1096}, country={Azerbaijan}}
\affiliation[label3]{organization={Department of Theoretical Physics, National University of Uzbekistan}, city={Tashkent}, postcode={100174}, country={Uzbekistan}}


\begin{abstract}
Measurement of numerical values of the anomalous density, $\sigma$, which plays important role in Bose -- Einstein condensation, and, especially, determination of  its sign, has been a long standing problem. We develop Hartree -- Fock -- Bogoliubov theory taking account arbitrary phase of the condensate wave function. We show that, the sign of $\sigma$ directly related to the phase, and, hence is not observable. Despite this, its absolute value can be extracted from measurements of the sound velocity and condensed fraction. We present  theoretical prediction for $\vert \sigma \vert$ for a BEC in a uniform box.
\end{abstract}
		
		
\end{frontmatter}
	
\section{Introduction}
\label{sec1}
In the systems with broken gauge symmetry, such as a system of bosons in very low temperatures, anomalous averages play an important role. At temperatures $T\le T_c$  in the dilute Bose gas   with critical temperature $T_c$, the absolute value of the anomalous average (density), $\vert\sigma\vert$ can be the same order as the normal particle density, $\rho_1$ \cite{yuk2005}.
Particularly, the importance of taking into account of the anomalous density
in mean - field theories has been revealed also in their application to
describe triplon condensation in quantum magnets. It has been shown that, neglecting
$\sigma $ leads to unreal jump in magnetization curves \cite{ourannals}.

Let $\psi(\textbf{r},t)$ is the field operator of bosonic particles.The spontaneous symmetry breaking (SSB) manifests itself via Bogoliubov shift \cite{bog47}
\begin{equation}
\psi(\textbf{r},t)=\chi+\tilde\psi(\textbf{r},t)\equiv e^{i\theta}\sqrt{\rho_0}+\tilde\psi(\textbf{r},t)
\label{shift}
\end{equation}  
where $\xi=\exp(i\theta)$ and $\rho_0$ are the phase and the density of condensed particles respectively, such that the average of the fluctuating field $\tilde\psi(\textbf{r},t)$ is zero $\langle\tilde\psi(\textbf{r},t)\rangle=0$. The normal and anomalous densities are defined as
\begin{equation}
\begin{array}{l}
\rho_1=\displaystyle\frac{1}{V}\displaystyle\int{\tilde\psi^{\dagger}}(\textbf{r})\tilde\psi(\textbf{r})d\textbf{r},\\
\\
\sigma=\displaystyle\frac{1}{V}\displaystyle\int\tilde\psi(\textbf{r})\tilde\psi(\textbf{r})d\textbf{r},
\label{defrho1sig}
\end{array}
\end{equation}
respectively. In the   case of a uniform system in the equilibrium the densities are constants. The physical meaning of the normal density  is straightforward: $\rho_1$ is the density of uncondensed particles with the normalization $\rho_0+\rho_1=\rho$, where $\rho=N/V$ is the density of total number of particles $N$. As to the physical meaning of $\sigma$, its absolute value, $\vert\sigma\vert$ gives the amplitude of  pair process, when two particles are annihilated from the thermal cloud of non - condensed particles. In  other words $\vert\sigma\vert$ describes the density of binary correlated particles.  Particularly, number of correlated pairs equals to $\vert\sigma\vert/2$.
		
What do we know about the sign of $\sigma$? One can find in the literature \cite{andersen,griffin}, the statement that $\sigma$ is negative i.e. $\sigma(T)<0$ or opposite to the sign of the interacting potential \cite{nozorez}. One of the goals of present work is making an attempt to find a more general answer to this question, assuming the phase of the condensed wave function $\xi$  to be arbitrary (clearly $\vert\xi\vert=1$).
		
The point is that in the literature the phase angle $\theta$ is set $\theta=0$ by default. This is realistic and natural for following reasons:
\begin{itemize}
\item In accordance with general rules of quantum mechanics, the physical observables must not depend on the phase of wave function;
\item The  phase and the amplitude, that is $\xi$ and $\rho_0$ in our notations are quantum mechanical conjugate quantities in the sense of Heisenberg uncertainty principle. When $\rho_0$ has been measured more or less exactly, $\xi$ remains unknown, and vice versa.
\item Let's consider interference of two Bose condensates with initially somehow fixed phases $\theta_1^0$ and $\theta_2^0$. Then contrary to our expectations the phase difference $\Delta\theta$ observed in the experiment would not be equal to to the initial difference i.e. $\Delta\theta \ne \theta_1^0-\theta_2^0$, since the mutual  interaction between the two condensates drastically changes the phase difference \cite{mogudno}. Actually this serves as a nice answer to the question ``Do superfluids that have never seen each other have a well-defined relative phase? In other words, does the BEC phase appears under the effect of SSB when it is formed, or later, when quantum measurement occurs?'', raised long years ago by Anderson \cite{anderson,mullin}.
\end{itemize}
		
On the other hand, the choice of the order parameter $\chi$ determines specific magnetic phases of spinor  Bose -- Einstein condensates \cite{kawagichi}. Moreover, it was shown \cite{ouraniz2p1,ouraniz2p2} that, in quantum magnets with weak anisotropies, where the direction of the magnetization vector is related to the phase of the triplon condensate \cite{giamarchi}, there may exist  some constraints to $\xi$. In present work we shall not discuss such magnetic effects, and concentrate mainly on the dependence of the sign of $\sigma$ on the phase  and its measurability for Bose - Einstein condensation (BEC) of atomic gases. We show that this dependence is rather simple, $\sigma\sim -\vert\sigma\vert\xi^2$, that is $\sigma$ may be negative or positive depending on the phase, which may vary randomly. As to its measurability, although BEC has been studied thoroughly in various laboratories, there is no report on the estimation of $\sigma$. We shall show that, in the framework of Hartree - Fock - Bogoliubov (HFB) theory $\vert\sigma\vert$ is 
 related to the sound velocity $s$ and the condensed fraction      $n_0=\rho_0/\rho$. Therefore it can be extracted from appropriate measurements of $s$ and $n_0$ at least for the dilute gases in a uniform box \cite{gaunt,navon}.

It is well known that, in high energy particle physics SSB leads to the emergence of Goldstone modes in
particle mass spectrum.  In quantum statistical mechanics , where BEC can take place, SSB imposes a condition to the excitation spectrum of the  system making it gapless. Namely, the energy dispersion of phonons (bogolons), $E_k$ should be linear in the long wavelength  limit, $E_k(k \to 0)\approx sk+{\cal O}(k^3)$, where $s$ is interpreted as a sound velocity (second sound in a superfluid system).  Long years ago Hugenholtz and Pines \cite{HP}, proved  that, this is equivalent to the following condition for the self energies:
\begin{equation}
\Sigma_n(0,0)-\Sigma_{an}(0,0)=\mu
\label{hp}
\end{equation}      
where $\mu$ is the chemical potential , and $\Sigma_n$ and $\Sigma_{an}$ are the normal and anomalous self energies, respectively. Unfortunately, the authors considered only the case with $\xi=1$, i.e. $\theta=0$ in Eq. \eqref{shift}. This relation has been recently extended for an arbitrary $\xi$ by Watabe \cite{watabehp}. It  is given by
\begin{equation}
\Sigma_n(0,0)-e^{-2i\theta}\Sigma_{an}(0,0)=\mu
\label{hpW}
\end{equation} 
In present work we show that Hugenholtz -- Pines -- Watabe (HPW) relation \eqref{hpW} together with the requirement of stability of the equilibrium system leads to the following constraint for the phase of the condensate wave function:
\begin{equation}
\xi^4=1, \quad i.e. \quad  \xi^2=\pm 1
\label{etatort1}
\end{equation}
On the first  glance, any constraint to the phase may confuse a reader.  However, one should  keep in mind that, due to the interactions and temperature fluctuations there always exists a depletion, and hence the general rule of quantum mechanics on the phase must be applied to the whole wave function $\psi=\exp(i\Theta)\vert\psi\vert$, given by \eqref{shift}. Clearly, the global phase of the latter $\Theta$ remains real and  fully arbitrary, due to the arbitrariness of the phase  of its fluctuating part $\tilde\psi$, which is usually integrated out in practical calculations.

In present work we perform calculations within HFB theory developed in Refs. \cite{andersen,griffin,tutorial,yukrepresent}. We extend this theory for an arbitrary $\xi$ and analyze its consequences for a homogeneous system of dilute Bose gases. For concreteness,  we follow the framework proposed by Yukalov \cite{yukrepresent}. This approach relies on introducing two different chemical potentials -- one for the condensed phase, and a second one for the normal phase. The two chemical potentials are distinct below the critical temperature, and allow us to simultaneously satisfy the self-consistency condition for the mean field while having a gapless Goldstone mode. They become equal at the critical temperature. Above the critical temperature, there is of course a single chemical potential, associated with the conserved particle number. Like the other aforementioned approaches to the Hohenberg -- Martin dilemma \cite{hohmartin}, Yukalovs prescription  introduces an additional variable to account for the different constraints. Note that this version of HFB approximation has been successfully applied  to describe experimental magnetization in quantum dimerized magnets \cite{ourannals,ourmce,ourcharak}, two component Bose gases \cite{ourbosemix,ourrabi}, and even to study quark matter in neutron stars \cite{hindiyuk} recently.

This work is organized as follows. In Sect. \ref{sec2} on the basis of extended HPW relation we derive a constraint to the phase of the condensed wave function for uniform system. In Sect. 3 we present a new formulation of HFB theory for an arbitrary phase $\theta$. Its phase
 invariance will be proven in the Sect. \ref{sec4}. In Sect. \ref{sec5} our numerical predictions for $\vert\sigma\vert$ are presented. And the last section includes our conclusions.

\section{Constraint to the phase of the condensate wave function}
\label{sec2}

The presence of anomalous density $\sigma$ in condensed phase ($T\le T_c$) leads to the puzzle, referred in the literature as Hohenberg -- Martin dilemma \cite{hohmartin}. The essence of this dilemma is following. In mean field theories the chemical potential may be determined in two alternative ways:
\begin{itemize}
	\item[a)] From the minimization of the grand thermodynamic potential $\Omega$ with respect to the order parameter, i.e. $\partial \Omega/\partial \chi$=0, which is, in some sense, equivalent to the condition $\langle H^{(1)}\rangle=0$ \cite{stoofbook}, 
	where $H^{(1)}$ is the linear part of the Hamiltonian in $\tilde\psi$.
	\item[b)] Or, simply from the Hugenholtz and Pines (HP) relation,   Eq.\eqref{hp}.
\end{itemize}

As a rule,  the difference between these chemical potentials does not vanish in the BEC phase, being proportional to the anomalous density, i.e. $\vert\mu_a-\mu_b\vert\propto \vert\sigma\vert$. This dilemma has been resolved by Kleinert and Yukalov \cite{klyuk}, in the concept of representative ensembles by introducing two Lagrange multipliers $\mu_0$ and $\mu_1$. In terms of these chemical potentials the standard grand Hamiltonian 
\begin{equation}
H[\psi^{\dagger},\psi]={\hat H}[\psi^{\dagger},\psi]-\mu {\hat N}[\psi^{\dagger},\psi] 
\label{Hgrand}
\end{equation}
should be replaced by 
\begin{equation}
H[\psi^{\dagger},\psi]={\hat H}[\psi^{\dagger},\psi]-\mu_0 {\hat N_0}[\psi^{\dagger},\psi]
-\mu_1 {\hat N_1}[\psi^{\dagger},\psi] 
\label{Hgrandyuk}
\end{equation}
such that thermodynamic relations with respect to the free energy $F$ and conserved numbers of particles, 
$\mu_0=\partial F/\partial N_0$ and $\mu_1=\partial F/\partial N_1$ are  satisfied. These two chemical potentials are no longer required to be equal. In fact, $\mu_0$ is mostly responsible for the conservation of particle number, while  $\mu_1$ for making the energy spectrum gapless via HP relation, which, for instance, for  $\xi^2=1$ case, includes $\mu_1$ instead of  $\mu=\mu_0n_0+\mu_1 n_1$:
\begin{equation}
\Sigma_n(0,0)-\Sigma_{an}(0,0)=\mu_1
\label{hpmu1}
\end{equation}
Here, before  discussing extended version of  this relation for arbitrary phase, we 
have to clarify definition of self energies.

\subsection{Self energies $X_1$ and $X_2$}
\label{subsec21}

The self energies are usually defined in terms of Green functions.
For the fluctuating complex fields, one introduces
a vector $(\tilde\psi,\tilde\psi^{\dagger})$ such that $G=i\langle{\cal T}(\tilde\psi,\tilde\psi^{\dagger})\cdot (\tilde\psi,\tilde\psi^{\dagger})^\dagger\rangle$,
that is \cite{Fetter,korealikkitob}
\begin{equation}
\begin{array}{l}
G_{11}(x,x')=-i\langle{\cal T}\tilde\psi(x)\tilde\psi^{\dagger}(x')\rangle\\
G_{12}(x,x')=-i\langle{\cal T}\tilde\psi(x)\tilde\psi(x')\rangle\\
G_{21}(x,x')=-i\langle{\cal T}\tilde\psi^{\dagger}(x)\tilde\psi^{\dagger}(x')\rangle\\
G_{22}(x,x')=-i\langle{\cal T}\tilde\psi^{\dagger}(x)\tilde\psi(x')\rangle
\label{Gt}
\end{array}
\end{equation}
Therefore, in general,  the operator $\widehat G$ has the form of a $2\times 2 $ matrix:
\begin{equation} 
	\widehat G=\left(
	\begin{matrix}
		{\widehat G}_{11} & {\widehat G}_{12}\\
		{\widehat G}_{21}&{\widehat G}_{22}
	\end{matrix}\right)
	\label{gmatr}
\end{equation}
Then ${\widehat\Sigma}$ is given by Dyson -- Beliaev equation \cite{stoofbook,abrosovkitob}
\begin{equation}
\widehat G^{-1}=\widehat G_{0}^{-1}-\widehat\Sigma
\label{dysonminusbir}
\end{equation}
where $\widehat G_{0}$ is the ``non - interacting'' Green function\footnote{In general $G_0$ may also include some part of the interaction, as it is always practiced in Optimized Perturbation Theory \cite{yukopt}.}. In most of cases $\Sigma_{11}=\Sigma_{22}$, $\Sigma_{12}=\Sigma_{21}$ and therefore $\Sigma_{11}$ and $\Sigma_{12}$ are referred as normal and anomalous self energies, respectively: $\Sigma_{n}\equiv\Sigma_{11}$, $\Sigma_{an}\equiv\Sigma_{12}$.

In practice these self energies can be extracted directly from the quadratic (bilinear) in fields part of the  Hamiltonian $H^{(2)}$ as follows. Let after the Bogoliubov shift \eqref{shift} the $H^{(2)}$ term acquires the form
\begin{equation}
H^{(2)}=\displaystyle\sum_k[\varepsilon_k-\mu_1+A]a^\dagger_ka_k+\displaystyle\frac{B}{2}
\displaystyle\sum_k[a^\dagger_{-k} a^\dagger_k+a_{-k}a_k]
\label{h2bill}
\end{equation}
where $a^\dagger_k$ ($a_k$) are creation (annihilation) operators introduced by
\begin{equation}
\begin{array}{l}
\tilde\psi(\textbf{r})=\displaystyle\frac{1}{\sqrt{V}}\displaystyle\sum_k e^{i\textbf{k}\textbf{r}}a_k,\quad
\tilde\psi^{\dagger}(\textbf{r})=\displaystyle\frac{1}{\sqrt{V}}\displaystyle\sum_k e^{-i\textbf{k}\textbf{r}}a^\dagger_k
\end{array}
\label{akakdag}
\end{equation}
Then one may conclude that \cite{stoofbook}
\begin{equation}
\Sigma_n=A, \quad \Sigma_{an}=B  
\label{sigAB}
\end{equation}

On the other hand it is more convenient  to present $\tilde\psi$ in terms of real fields
$\psi_1$  and $\psi_2$, as \cite{andersen}
\begin{equation}
\begin{array}{l}
\tilde\psi=\displaystyle\frac{1}{\sqrt{2}}(\psi_1+i\psi_2)\\
\tilde\psi^\dagger=\displaystyle\frac{1}{\sqrt{2}}(\psi_1-i\psi_2)
\end{array}
\label{psi12}
\end{equation}
with the Green function $D_{ij}=\langle T_\tau \psi_i \psi_j\rangle$, $(i,j=1,2)$. It can be shown that \cite{andersen,ourctan,mysolo}, in Matsubara formalism
\begin{equation}
\label{psip}	
\begin{array}{l}
\psi_i(\tau,\textbf{r})=\displaystyle\frac{1}{\sqrt{V\beta}}\displaystyle\sum_k
\displaystyle\sum_{n=-\infty}^{\infty}\psi_i(\omega_n,\textbf{k})
e^{i\omega_n\tau+i\textbf{k}\textbf{r}}, \\
D_{ij}(\tau,\textbf{r};\tau',\textbf{r}')=\displaystyle\frac{1}{{V\beta}}\displaystyle\sum_k
\displaystyle\sum_{n=-\infty}^{\infty}
e^{i\omega_n(\tau-\tau')}e^{i\textbf{k}(\textbf{r}-\textbf{r}')}D_{ij}(\omega_n,\textbf{k})
\end{array}
\end{equation}
the inverse Green function $D^{-1}$ has the  following compact form
\begin{eqnarray}
{D}^{-1}(\omega_n,\textbf{k})=\left(
\begin{matrix}
\varepsilon_k+X_1 & \omega_n\\
-\omega_n&\varepsilon_k+X_2
\end{matrix}
\right)
\label{Dinv}
\end{eqnarray}
where $\varepsilon_k=\textbf{k}^2/2m$ and $ \omega_n=2\pi n T$  is the Matsubara frequency $(\beta=1/T)$. Note that, above  we have  assumed the translation invariance, $D(x,x')=D(x-x')$, which holds for a homogeneous system. The self energies $X_{1,2}$ are related to the ordinary ones as follows:
\begin{equation}
\Sigma_n=\mu_1+\displaystyle\frac{X_1+X_2}{2},\quad \Sigma_{an}=\displaystyle\frac{X_1-X_2}{2}
\label{x12sig}
\end{equation}
As to the energy  dispersion, $E_k$ corresponds to the zeros of the inverse propagator: $Det[D^{-1}]=0$ and,
clearly,  has the form
\begin{equation}
E_k=\sqrt{(\varepsilon_k+X_1)(\varepsilon_k+X_2)}
\label{Ekx1x2}
\end{equation}

\subsection{HPW relation}
\label{subsec22}

In the formalism with two chemical potentials the Hugenholtz -- Pines -- Watabe relation 
is rewritten as
\begin{equation}
\Sigma_n(0,0)-{\bar\xi}\,^2\Sigma_{an}(0,0)=\mu_1
\label{hpWmu1}
\end{equation}
where ${\bar\xi}=\xi^+=\exp(-i\theta)$. That is $\mu$ in \eqref{hpW} should be simply 
 replaced by $\mu_1$.
This acquires more symmetric form in terms of $X_{1,2}$. In fact, insertnig
\eqref{x12sig} into \eqref{hpWmu1}  one obtains
\begin{equation}
X_1(1-\xi^2)-X_2(1+\xi^2)=0
\label{hpwx12}
\end{equation}

Now we prove the constraint $\xi^2=\pm 1$ for a stable uniform  BEC in the equilibrium. Actually, it is clear that, the dispersion $E_k$ in \eqref{Ekx1x2} , as well as  the variables $X_{1,2}$ must be real and positive in a stable BEC phase, i.e. $X_{1,2}\ge 0$. Presenting $\xi^2$ as 
\begin{equation}
\xi^2=e^{(2i\theta)}=\cos 2\theta+i\sin 2\theta 
\label{xisincos}
\end{equation}
and separating real and imaginary parts of Eq. \eqref{hpwx12} we get
\begin{equation}
X_1(1-\cos 2\theta )-X_2(1+\cos 2\theta )=0,
\label{etareal}
\end{equation}
\begin{equation}
(X_1+X_2)\sin 2\theta=0.
\label{etarealim}
\end{equation}
The equation \eqref{etarealim} has two solutions: $X_1+X_2=0$, and (or) $\sin 2\theta=0$. It is understood that the first solution $X_1=-X_2$ is  nonphysical.
 In fact, in this case both $X_1$ and $X_2$ can not be positive simultaneously. Moreover, if $X_1=-X_2=0$ then for the dispersion we would have $E_k=\varepsilon_k=\textbf{k}^2/2m$, which contradicts to Goldstone theorem. Therefore, the only solution 
\begin{equation}
\sin 2\theta=0, \quad i.e. \quad  \cos 2\theta=\pm 1=\xi^2
\label{sincos}
\end{equation}
may be chosen as a physical one. Therefore, inserting \eqref{sincos} into \eqref{etareal} 
one can conclude that either $X_1$ or $X_2$ must be equal to zero. For example, if
$\xi^2=1$ then $X_2=0$, and the dispersion $E_k=\sqrt{\varepsilon_k(\varepsilon_k+X_1)}$ will be gapless.
Thus , we have proven that HPW relation \eqref{hpwx12} together with the stability condition
in BEC phase imposes following condition to the phase:${\bar \xi}^2=\xi^2=\pm 1$, i.e.
$\xi=\pm 1, \pm i$. It should be underlined that, the constraint concerning the order parameter
does not violate general laws of quantum mechanics on the phase of wave function. Actually,
the whole wave function of the system is described by $\psi=\xi \sqrt{\rho_0}+\tilde\psi\equiv \exp(i\Theta)\vert\psi\vert$, where the phase $\Theta$ can take any real value, due to the presence of complex field $\tilde\psi$. However, the phase angle of the condensate wave function acquires
randomly only discrete values: $\theta=\pi n/2$, ($n=0.\pm 1, \pm 2 \dots$). 

It is understood that the Hamiltonian of the Bose system is gauge invariant, i.e. does not depend on $\Theta$. As to the invariance of thermodynamic quantities in HFB with respect to $\theta$, it will be demonstrated in Sec. \ref{sec4}. Meanwhile, we have to fix the Hamiltonian and diagonalize it to find explicit expressions for observables. Below, in general, we use the units $k_B=1$ and $\hbar=1$.

\section{HFB theory for an arbitrary $\xi$}
\label{sec3}

We start with  the standard Hamiltonian for homogeneous system with contact interaction
\begin{equation}
{\hat H}=\displaystyle\int d\textbf{r}\displaystyle\frac{\nabla\psi^\dagger(\textbf{r}) \nabla \psi(\textbf{r})}{2m}+ \displaystyle\frac{g}{2}\displaystyle\int d\textbf{r}\left[\psi^\dagger(\textbf{r}) \psi (\textbf{r}) \right]^2
\label{hamham}
\end{equation}
where coupling constant $g$ is related to the s-wave scattering length $a_s$   via $g=4\pi a_s/m$. Implementing here Bogoliubov shift \eqref{shift} one splits the grand Hamiltonian \eqref{Hgrandyuk} as follows:
\begin{eqnarray}
H&=&H^{(0)}+H^{(1)}+H^{(2)}+H^{(3)}+H^{(4)}\nonumber\\
H^{(0)}&=&V(-\mu_0\rho_0+\displaystyle\frac{g\rho_0^2}{2})\nonumber\\
H^{(1)}&=&\sqrt{\rho_0}(g\rho_0-\mu_0)\displaystyle\int d\textbf{r}({\bar\xi}\tilde\psi(\textbf{r})+h.c)
\label{hspar}\\
H^{(2)}&=&\displaystyle\int d\textbf{r}\left\{
\tilde\psi^\dagger(\textbf{r}) \left[-\displaystyle\frac{{\bf\nabla}^2}{2m} -\mu_1 +2g\rho_0
\right]\psi(\textbf{r})+\displaystyle\frac{g\rho_0\xi^2}{2}\left[
\tilde\psi^2(\textbf{r})+\tilde\psi^{\dagger\,2}(\textbf{r})
\right]\right\}\nonumber\\
H^{(3)}&=&g\sqrt{\rho_0}\displaystyle\int d\textbf{r}\tilde\psi^\dagger(\textbf{r})\tilde\psi(\textbf{r})
\left[\tilde\psi^\dagger(\textbf{r})\xi+h.c.\right]\nonumber\\
H^{(4)}&=&\displaystyle\frac{g}{2}\displaystyle\int d\textbf{r} \left[\tilde\psi^\dagger(\textbf{r})\tilde\psi(\textbf{r})\right]^2
\nonumber
\end{eqnarray}
where we have taken into account ${\bar\xi}^2=\xi^2$. In momentum space using Eqs. \eqref{akakdag} these can be rewritten as 
\begin{equation}
\begin{array}{l}
H^{(1)}=\displaystyle\frac{\sqrt{\rho_0}}{\sqrt{V}}\displaystyle\sum_k\left[a^\dagger_k\xi(g\rho_0-\mu_0)+h.c.
\right]\\
H^{(2)}=\displaystyle\sum_k \left[\varepsilon_k-\mu_1+2g\rho_0\right] a^\dagger_k a_k+
\displaystyle\frac{\xi^2 g\rho_0}{2}\displaystyle\sum_k \left[a_ka_{-k}+a^\dagger_ka^\dagger_{-k}\right]\\
H^{(3)}=g\sqrt{\rho_0}\displaystyle\sum_{k,p}\left[\xi{a}_p^\dagger a_{p-k}a_k +h.c.\right]\\
H^{(4)}=\displaystyle\frac{g}{2V}\displaystyle\sum_{k,p,q}a^\dagger_k a_p^{\dagger}a_q a_{k+p-q} 
\end{array}
\label{hakka}
\end{equation}
To treat cubic and quartic terms we implement following substitutions, based on the Wick's theorem:
\begin{equation}
\begin{array}{lcl}
a_{\mathbf k}^{\dagger}a_{\mathbf p}a_{\mathbf q}&\rightarrow&
2\langle a_{\mathbf k}^{\dagger} a_{\mathbf p}\rangle a_{\mathbf q}+a_{\mathbf k}^{\dagger}\langle a_{\mathbf p} a_{\mathbf p}\rangle\\
a_{k}^{\dagger}a_{p}^{\dagger}a_{q}a_{m}&\rightarrow& 
4a_{k}^{\dagger}a_{m}\langle a_{p}^{\dagger}a_{q}\rangle+a_{q}a_{m}\langle a_{k}^{\dagger}a_{p}^{\dagger}\rangle
+a_{k}^{\dagger}a_{p}^{\dagger}\langle a_{q}a_{m}\rangle -2\rho_{1}^{2}-\sigma^{2}
\end{array}	
\label{wikss}
\end{equation}
where  $\rho_{1}$ and  $\sigma $ are normal and anomalous densities, defined in Eq. \eqref{defrho1sig},  have the form 
\begin{eqnarray}
\rho_{1}&=&\displaystyle\frac{1}{V}\displaystyle\sum_{k}n_{k}=\displaystyle\frac{1}{V}
\displaystyle\sum_{k} \langle a_{k}^{\dagger}a_{k}\rangle 
\label{rho1ak}\\
\sigma&=&\displaystyle\frac{1}{V}\displaystyle\sum_{k}\sigma_{k}=\displaystyle\frac{1}{2V}\displaystyle\sum_{k} \left[\langle a_{k}a_{-k}\rangle + \langle a_{k}^\dagger a_{-k}^\dagger\rangle\right]
\label{sigmaak}
\end{eqnarray}
where we used following properties of bosonic  operators: $\langle a_{k}^{\dagger}a_{p}\rangle=\delta(\textbf{k}-\textbf{p})n_{k}$, $\langle a_{k}a_{p}\rangle = \delta(\textbf{k}+\textbf{p})\sigma_{k}$.

Then $H^{(3)}$ and $H^{(4)}$ will contribute to the  other terms, resulting in
\begin{equation}
\begin{array}{l}
H=H^{Class}+H^{Lin}+H^{Bilin}\\
H^{Class}=V[-\mu_0\rho_0+\displaystyle\frac{g\rho_0^2}{2}]-\displaystyle\frac{g}{2V}
\left[2\rho_1^2+\sigma^2\right]\\
H^{Lin}=\sqrt{\rho_0}\displaystyle\sum_k \left\{\left[g\sigma\xi+(2g\rho-g\rho_0-\mu_0){\bar\xi}\right]a^\dagger_k+h.c.\right\}\\
H^{Bilin}=\displaystyle\sum_k \left\{\left[\varepsilon_k-\mu_1+2g\rho\right] 
a^\dagger_k a_k+\displaystyle\frac{g(\sigma+\xi^2\rho_0)}{2}
\left[a_k a_{-k}+a_k^\dagger a_{-k}^\dagger\right]\right\}
\end{array}
\label{Hamakaka}
\end{equation}
The average of the linear part allows to fix $\mu_0$. So, from the condition $\langle H^{Lin}\rangle=0$ one immediately obtains:
\begin{equation}
\mu_0=g\sigma\xi^2+2g\rho-g\rho_0
\label{mu00}
\end{equation}    
The bilinear part, in accordance with the rules   \eqref{h2bill} -- \eqref{sigAB}, gives rise to the self
energies as
\begin{equation}
\Sigma_n=2g\rho, \quad \Sigma_{an}=g(\sigma+\rho_0\xi^2)=
g[\sigma+(\rho-\rho_1)\xi^2]
\label{sigabrule}
\end{equation}
which lead to following main equations with respect to $X_{1}$  and $X_2$
\begin{eqnarray}
X_1=-\mu_1+2g\rho+g\sigma(X_1,X_2)+g\rho_0(X_1,X_2)\xi^2=\Sigma_n+\Sigma_{an}-\mu_1
\label{mainx1eq}\\
X_2=-\mu_1+2g\rho-g\sigma(X_1,X_2)-g\rho_0(X_1,X_2)\xi^2=\Sigma_n-\Sigma_{an}-\mu_1
\label{mainx2eq}
\end{eqnarray}
with the normalization $\rho_0(X_1,X_2)=\rho-\rho_1(X_1,X_2)$. They  remain as formal equations  unless one finds  functions $\sigma(X_1,X_2)$ and
$\rho_1(X_1,X_2)$ explicitly. This procedure can be performed by using Bogoliubov transformations: 
\begin{equation}
a_{\mathbf k}=u_{\mathbf k}b_{\mathbf k}+v_{\mathbf k}b_{-{\mathbf k}}^{\dagger},\quad 
a_{\mathbf k}^{\dagger}=u_{\mathbf k}b_{\mathbf k}^{\dagger}+v_{\mathbf k}b_{-{\mathbf k}}
\label{Bogtrans}
\end{equation}
The operators $b_{\mathbf k}$ and $b_{\mathbf k}^{\dagger}$ can be interpreted
as annihilation and creation Bose operators of phonons (bogolons) with following properties.
\begin{equation}
\begin{array}{l}
[b_{\mathbf k},b_{\mathbf p}^{\dagger}]=\delta_{\mathbf{k},\mathbf{p}},\quad  \langle b_{\mathbf k}^{\dagger}b_{-{\mathbf k}}^{\dagger}\rangle=\langle b_{\mathbf k}b_{-\mathbf{k}}\rangle=0,\\
\langle b_{\mathbf k}^{\dagger}b_{\mathbf k}\rangle=
\displaystyle\frac{1}{e^{\beta E_{\mathbf k}}-1}\equiv f_{B}(E_{\mathbf k})
\end{array}
\label{bbkbkk}
\end{equation}
where $E_k$ is formally given by Eq. \eqref{Ekx1x2}. 

To justify  the relation \eqref{Ekx1x2} we insert \eqref{Bogtrans} into $H^{Bilin}$ and require that the coefficient
of the term $(b_{\mathbf k}b_{-\mathbf{k}}+b_{-\mathbf{k}}^{\dagger}b_{\mathbf k}^{\dagger})$ vanishes, i. e:
\begin{equation}
\omega_{\mathbf k}u_{\mathbf k}v_{\mathbf k}+\displaystyle\frac{\Sigma_{an}}{2}(u^{2}_{\mathbf k}+
v_{\mathbf k}^{2})=0.
\label{condforb}
\end{equation}
where $\omega_k=\varepsilon_k-\mu_1+\Sigma_n$. Now using the condition $u_{\mathbf k}^{2}-v_{\mathbf k}^{2}=1$ and presenting $u_{\mathbf k}, v_{\mathbf k}$ as
\begin{equation}
u_{\mathbf k}^{2}=\frac{\omega_{\mathbf k}+E_{\mathbf k}}{2E_{\mathbf k}},
\quad v_{\mathbf k}^{2}=\frac{\omega_{\mathbf k}-E_{\mathbf k}}{2E_{\mathbf k}}
\label{ukvkEk}
\end{equation}
yields
\begin{equation}
\sqrt{\omega_{\mathbf k}^{2}-E_{\mathbf k}^{2}}=-\Sigma_{an},\quad
u_k v_k=-\displaystyle\frac{\Sigma_{an}}{2E_k},\quad
u_{\mathbf k}^{2}+v_{\mathbf k}^{2}=\displaystyle\frac{\omega_k}{E_k}
\label{promij}
\end{equation}
that is $E_k^2=(\omega_k+\Sigma_{an})(\omega_k-\Sigma_{an})=(\varepsilon_k+X_1)(\varepsilon_k+X_2)$,
where $X_{1,2}$ are the solutions of the Eqs. \eqref{mainx1eq}, \eqref{mainx2eq}. For the densities
\eqref{rho1ak} and \eqref{sigmaak} one obtains
\begin{equation}
\begin{array}{l}
\rho_1=\displaystyle\frac{1}{V}\displaystyle\sum_k\langle a^\dagger_k a_k\rangle =
\displaystyle\frac{1}{V}\displaystyle\sum_k\left[f_B(E_k)(u_k^2+v_k^2)+v_k^2\right]\\
\sigma=\displaystyle\frac{1}{V}\displaystyle\sum_k\langle a_k a_{-k}\rangle =
\displaystyle\frac{1}{V}\displaystyle\sum_k u_kv_k \left[1+2f_B(E_k)\right]
\end{array}
\label{densr1s}
\end{equation}
These will take more elegant form in terms of $X_{1,2}$:
\begin{eqnarray}
\rho_1&=&\displaystyle\frac{1}{V}\displaystyle\sum_k 
\left\{\displaystyle\frac{W_k\left[\varepsilon_k+(X_1+X_2)/2\right]}{E_k}-
\displaystyle\frac{1}{2}\right\}
\label{rho1x12}\\
\sigma&=&\displaystyle\frac{X_2-X_1}{2V}\displaystyle\sum_k
\displaystyle\frac{W_k}{E_k}
\label{sigmasx12}
\end{eqnarray}
where we used Eqs.  \eqref{mainx1eq}, \eqref{mainx2eq}, \eqref{promij}, \eqref{densr1s} and introduced the notation
\begin{equation}
W_k=\displaystyle\frac{1}{2}+f_B(E_k)=\displaystyle\frac{1}{2}\coth(E_k/2T)
\label{Wk}
\end{equation}

Now due to Eqs. \eqref{condforb} -- \eqref{promij} $H^{Bil}$ in  the grand Hamiltonian \eqref{Hamakaka} is diagonal. Therefore, the total energy per volume of the system can be found from $E/V=(H+\mu_0N_0+\mu_1N_1)/V$ which has following explicit form:
\begin{equation}
\begin{array}{lcl}	
\displaystyle\frac{E}{V}&=&-\displaystyle\frac{X_1(\sigma+\rho_1)}{2}+\displaystyle\frac{X_2(\sigma-\rho_1)}{2}+
g\rho_1^2+2g\rho_1\rho_0+\xi^2 g\rho_0\sigma+\\
&&+\displaystyle\frac{g(\sigma^2+\rho_0^2)}{2}+\displaystyle\frac{1}{V}E_{LHY}+\frac{1}{V}\sum_k E_k f_B(E_k)
\end{array}
\label{ETotV}
\end{equation}
In  \eqref{ETotV} the zero  mode (Lee -- Huang -- Yang) term is given by
\begin{equation}
E_{LHY}=\frac{1}{2}\displaystyle\sum_k \left[E_k-\varepsilon_k-\frac{X_1+X_2}{2}+
\frac{(X_1-X_2)^2}{8\varepsilon_k}\right]
\label{Elhy}
\end{equation}
where we included counter terms to make finite the integration over momentum.
One may also calculate the thermodynamic free energy to obtain 
\begin{equation}
F=E+T\displaystyle\sum_k\left[\ln(1-e^{-\beta E_k})-\beta E_k f_B(E_k)\right]
\label{Ftot}
\end{equation}
It can be also shown that, minimization of the grand thermodynamic potential $\Omega=F-\mu_0 N_0 - \mu_1 N_1$
with respect to $\rho_0$ leads to the same expression  for $\mu_0 $ as the Eq. \eqref{mu00}, which
is in  favor of the self-consistency of the present theory. On the other hand, it is seen that,the thermodynamic potentials in \eqref{ETotV} -- \eqref{Ftot} include the phase  $\xi^2=\exp(2i\theta)$ explicitly or via $X_{1,2}(\xi)$. In the next section we show that this dependence is, in  fact, only apparent
.

\section{Phase invariance of physical quantities in HFB}
\label{sec4}

Since the self energies $X_{1,2}$ given by Eqs. \eqref{mainx1eq} and \eqref{mainx2eq} explicitly depend on $\xi^2$, the densities $\rho_1(X_1,X_2)$ and $\sigma(X_1,X_2)$, contributing to the thermodynamic potentials seem also to be  phase dependent.  On the other hand, in   Sect.2 we have proven that the phase of the condensate wave function can take values only $\xi=\pm 1, \;\; \xi=\pm i$, that is $\xi^2=1$ or $\xi^2=-1.$ Obviously, a physical quantity should not depend on the sign of $\xi^2$, in other words it must be invariant under the transformation ${\widehat P}_\xi$ defined as $\widehat P_\xi f(\xi^2)= f(-\xi^2)$.

To facilitate further considerations we make an attempt to extract explicit dependence of $\sigma$ on $\xi^2$. Firstly, inserting $X_{1,2}$ from Eqs. \eqref{mainx1eq},
\eqref{mainx2eq} into \eqref{sigmasx12} we obtain 
\begin{equation}
\sigma=\displaystyle\frac{X_2-X_1}{2}\bar{S}
=-g\bar{S}(\sigma+\xi^2\rho_0) 
\label{sigeq}
\end{equation}
whose formal solution is
\begin{equation}
\sigma=-\displaystyle\frac{g\bar{S} \xi^2\rho_0}{1+g\bar{S}}\equiv -\xi^2\bar{\sigma}
\label{sigeqq}
\end{equation}
where we introduced following notations
\begin{equation}
\bar{S}=\displaystyle\frac{1}{V}\displaystyle\sum_k\displaystyle\frac{W_k}{E_k},\quad 
\bar{\sigma}=\displaystyle\frac{g\bar{S} \rho_0}{1+g\bar{S}}
\label{sigeqqnot}
\end{equation}
From Eq. \eqref{sigeqq} it is seen that, $\sigma$ vanishes in the normal phase together
with $\rho_0$, i.e. $\sigma(T\ge T_c)=\rho_0(T\ge T_c)=0$. That is there are
no binary correlated particles out of the condensate phase.

Further, inserting \eqref{sigeqq} back to Eqs. \eqref{mainx1eq} and \eqref{mainx2eq}
we rewrite the latters  as
\begin{equation}
X_1=-\mu_1+2g\rho+\xi^2 g (\rho_0-\bar{\sigma})
\label{x1sigbar}
\end{equation}
\begin{equation}
X_2=-\mu_1+2g\rho-\xi^2 g (\rho_0-\bar{\sigma})
\label{x2sigbar}
\end{equation}
and putting these equations together, we obtain
\begin{equation}
\mu_1=2g\rho-\displaystyle\frac{X_1+X_2}{2}
\label{mu1eq}
\end{equation}

Now we are on the stage of discussing the properties of  each quantity under the transformation $\xi^2\to -\xi^2$separately.
\begin{itemize}
\item 
For the beginning,  it is  directly seen from Eqs. \eqref{x1sigbar}, \eqref{x2sigbar} that,
\begin{equation}
\widehat P_\xi X_1=X_2, \quad \widehat P_\xi X_2=X_1 
\label{hatpx12}
\end{equation}
that is under the transformation $\xi^2\to -\xi^2$ these variables will be reversed,
for instance, $X_1(\xi^2=+1)=X_2(\xi^2=-1)$ . Therefore, it is understood that, the dispersion $E_k$, the normal 
density $\rho_1$,   chemical potential $\mu_1$, and $\Sigma_n$ given by Eq.s
\eqref{Ekx1x2}, \eqref{rho1x12}, \eqref{mu1eq}  and \eqref{x12sig} respectively, are phase invariant, being symmetric in variables $X_{1,2}$. 

\item 
The main factor in the anomalous density, $\bar{\sigma}$, is symmetric i.e. $\bar{\sigma}(X_1,X_2)$ $=\bar{\sigma}(X_2,X_1)$, as it is seen from Eq. \eqref{sigeqqnot}, and hence $\widehat P_\xi \bar{\sigma}=\bar{\sigma}$. However, $\widehat P_\xi \sigma(\xi^2)=\sigma(-\xi^2)=-(-\xi^2)\bar{\sigma}=-\sigma$ that is in contrast to the normal density , anomalous density is phase dependent, having opposite signs in the states with the phases $\xi^2=+1$, and $\xi^2=-1.$ The same is true for $\Sigma_{an}$, which is antisymmetric in $X_{1,2}$  (see Eq. \eqref{x12sig}).

\item The total energy $E$. In equation \eqref{ETotV} the first two terms are symmetric under the transformation $X_1\leftrightarrow X_2$ by themselves. As to other terms they are also 
phase invariant. For example
\begin{equation}
\widehat P_\xi\sigma^2=(-\sigma)^2,\quad \widehat P_\xi(\xi^2\sigma)=(-\xi)^2(-\sigma)=\xi^2 \sigma 
\label{peee}
\end{equation}
Therefore, $E$ as well as the free energy $F$ are phase independent.
\end{itemize}

Thus we have proven in details that, in present version of HFB  physical observables do not depend on the phase of the condensate wave function, as expected. As to the anomalous density $\sigma$ we have shown that its sign is determined by $\xi^2=\exp(2i\theta)$, and hence could  not be observed
in reality. However, its absolute value, $\vert\sigma\vert$ could be observed and even measured. Below we discuss such procedure.

\subsection{How to measure $\vert\sigma\vert$?}
\label{subsec41}

Above we have proven that,  in actual calculations it does not matter which value of $\xi^2=\pm 1$ will be chosen. Particularly, if one chooses the case $\xi^2=+1$, then HPW relations \eqref{hpwx12} lead to $X_2=0$, while  the case with  $\xi^2=-1$ imposes $X_1=0$. In each case one is left, actually, with the same equations for physical quantities.

Let's for concreteness fix the  choice $\xi^2=1$. Then the main Eqs. \eqref{x1sigbar} and \eqref{x2sigbar} have the form
\begin{eqnarray}
X_1=-\mu_1+2g\rho+g(\rho-\rho_1-\bar{\sigma}),\label{mainx1eqsig}\\
X_2=-\mu_1+2g\rho-g(\rho-\rho_1-\bar{\sigma})\label{mainx2eqsig}
\end{eqnarray}
with $\bar{\sigma}$ defined in \eqref{sigeqqnot}. In accordance with \eqref{hpwx12} we set here $X_2=0$ to obtain
\begin{equation}
\mu_1=2g\rho-g(\rho-\rho_1-\bar{\sigma})
\label{mu1form}
\end{equation}
Inserting this to \eqref{mainx1eqsig} leads to the equation
\begin{equation}
\Delta\equiv \frac{X_1}{2}=
g(\rho_0+\sigma)=g(\rho-\rho_1-\bar{\sigma})=g\rho(n_0-{\bar m}_1)
\label{deleq}
\end{equation}
where
\begin{equation}
\begin{array}{l}
n_0=\displaystyle\frac{N_0}{N}=\displaystyle\frac{\rho_0}{\rho}=1-n_1, \quad {\bar m}_1=\bar{\sigma}/\rho\\
n_1=\displaystyle\frac{1}{V\rho }\displaystyle\sum_k\left[
\displaystyle\frac{W_k(\varepsilon_k+\Delta)}{E_k}-\displaystyle\frac{1}{2}
\right]
\end{array}
\label{settteq}
\end{equation}
with the dispersion 
\begin{equation}
E_k=\sqrt{\varepsilon_k}\sqrt{\varepsilon_k+2\Delta}
\label{dispdel}
\end{equation}
In practice the set of nonlinear algebraic  equations \eqref{deleq} -- \eqref{dispdel} can be easily solved numerically with respect to $\Delta$, being expressed in terms of only two input parameters, $\gamma=\rho a_s^3$ and $t=T/T_c$, as it is illustrated in the Appendix.

Considering the above we propose following strategy to estimate experimental values of $\vert\sigma\vert$ using the data on the condensed fraction $n_0$ and the sound velocity $s$. Thus,  for each set of $\gamma$ and $t$
\begin{enumerate}
\item 
Evaluate $\Delta(\gamma,t)$ from the experimental data on the sound velocity  $s$, by using the equation $\Delta=m s^2(\gamma,t)$;
\item
 Find $\vert\sigma\vert$ from the equation
\begin{equation}
m_1\equiv \frac{\vert\sigma\vert}{\rho}=n_0-\frac{\Delta}{g\rho}=
n_0-\frac{m s^2}{g\rho}
\label{m1sig}
\end{equation}
where  $\Delta(\gamma,t) $ has been  already fixed, and $n_0(\gamma,t)$ is given by
experiments.
\end{enumerate}

Unfortunately, present time,  this strategy  have not been  realized for following reasons:
\begin{enumerate} 
\item Strictly speaking, present theory is valid only for homogeneous dilute Bose gases with constant density, $\rho=const$;  

\item Although inhomogeneous BECs in magnetic traps have been well studied inside and out, the data on homogeneous condensates is quite scarce.   
\end{enumerate} 
In fact, we failed to find a report on experimental measurements 
of the condensate fraction $n_0$ and the sound velocity $s$ on the same material     
in a uniform box. Although, the first homogeneous BEC in optical box trap was created
twenty  years ago \cite{meyrat}, only the works  \cite{gaunt,lopes}, provide with
the experimental data on the condensate fraction, but not on the sound velocity.
In the next section we make an attempt to describe this data within present theory and
bring our predictions for $\vert\sigma\vert$.

\section{Numerical results and discussions}
\label{sec5}	

In 2013 experimenters in Cavendish Laboratory \cite{gaunt}, observed the BEC of atomic  gas of $^{87}Rb$ in the state $\vert F,m_F\rangle = \vert 2,2\rangle$ with constant density $\rho\approx 0.11\times 10^{14}/cm^3$. Bearing in mind the experimental value of the s- wave scattering length $a_s=5nm$, obtained by Brazilian group \cite{brazil}, we have $\gamma=\rho a_s^3\approx 1.4\times 10^{-6}$. It is understood that, for such a small value of the gas parameter, the critical temperature $T_c$ can be evaluated from the simple formula \cite{pitaevskibook}:
\begin{equation}
T_c=\displaystyle\frac{3.3125 \rho^{2/3}(\hbar c)^2
}{k_B mc^2}
\label{tc}
\end{equation}
with $k_B=8.617 \times 10^{-5} ev/K$, $(\hbar c)\approx 197.33\; nm\; ev$ and $mc^2\approx 87 \times 939\; Mev$. Actually, for $\rho\approx 0.11\times 10^{-7}/nm^3$ we get $T_c=92.17\,nK$ , which coincides with the experimental one with a great accuracy.

\begin{figure}[t]
\begin{minipage}[h]{0.48\linewidth}
\center{\includegraphics[width=1.\linewidth]{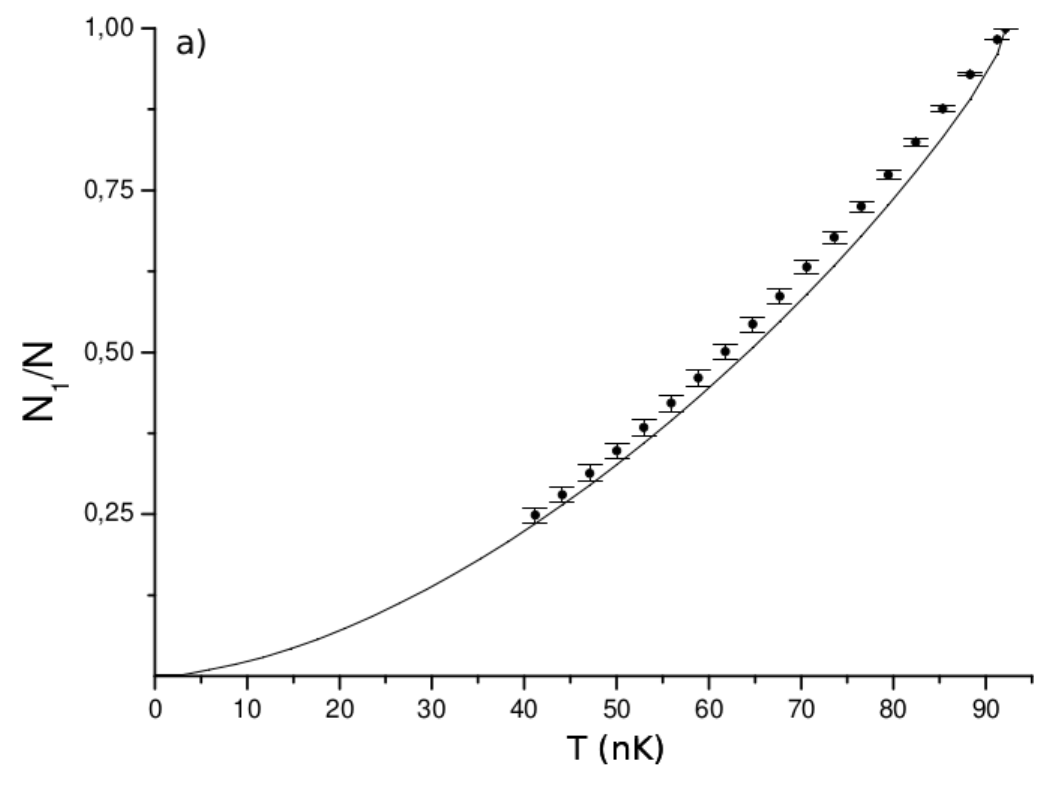}}
\end{minipage}
\hfill
\begin{minipage}[h]{0.48\linewidth}
\center{\includegraphics[width=0.95\linewidth]{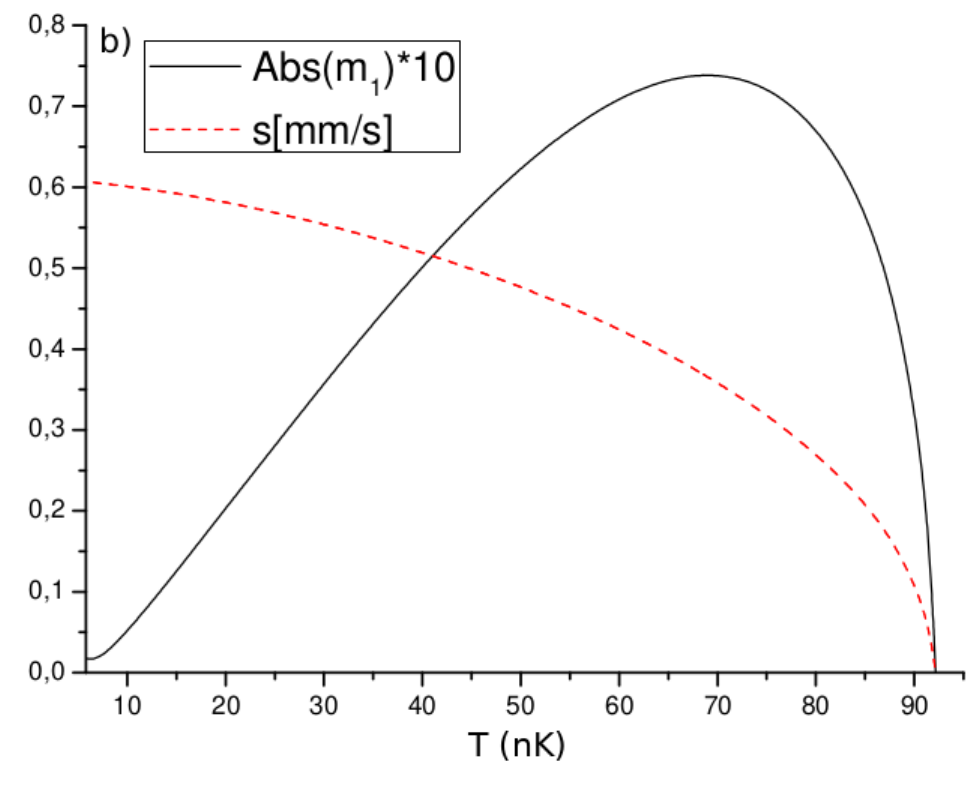}}
\end{minipage}
\hfill
\caption{(a) Fraction of uncondensed particles, $n_1=1-n_0$ vs temperature. Note that, at $T=0$ non-condensed fraction is very small: $n_1(T=0)=0.18\times 10^{-2}$. Experimental points are taken from  fitting experimental data in \cite{gaunt}. (b) Our prediction for the absolute value of the anomalous density, $\vert m_1\vert = \vert\sigma\vert/\rho$ (solid line) and sound velocity (dashed line) for  $^{87}Rb$ condensate in uniform box potential.}
\label{Fig1}
\end{figure} 

Using above parameters,  we have calculated $n_0$, $\Delta$, $s$ and $\vert\sigma\vert$ within present HFB theory. As it is seen from Fig.\ref{Fig1}(a), the theory describes existing data on $n_1=1-n_0$ nicely. It's nice to note that, there was no need to optimize input parameters.

 In Fig.\ref{Fig1}(b) we present our predictions for the sound velocity and the absolute value of the anomalous density vs  temperature for homogeneous BEC condensate in $^{87}Rb$ dilute gas. It is seen that, in contrast to the sound velocity, $\vert\sigma\vert$ does not decreases monotonically vs temperature. Actually, it first increases, reaches its maximum near $t\sim 0.7$ and then rapidly decreases to vanish  at $t=1$. Hopefully, such behavior of the anomalous density will be revealed also in future experiments.

\begin{figure}[t]
\begin{minipage}[h]{0.48\linewidth}
\center{\includegraphics[width=0.95\linewidth]{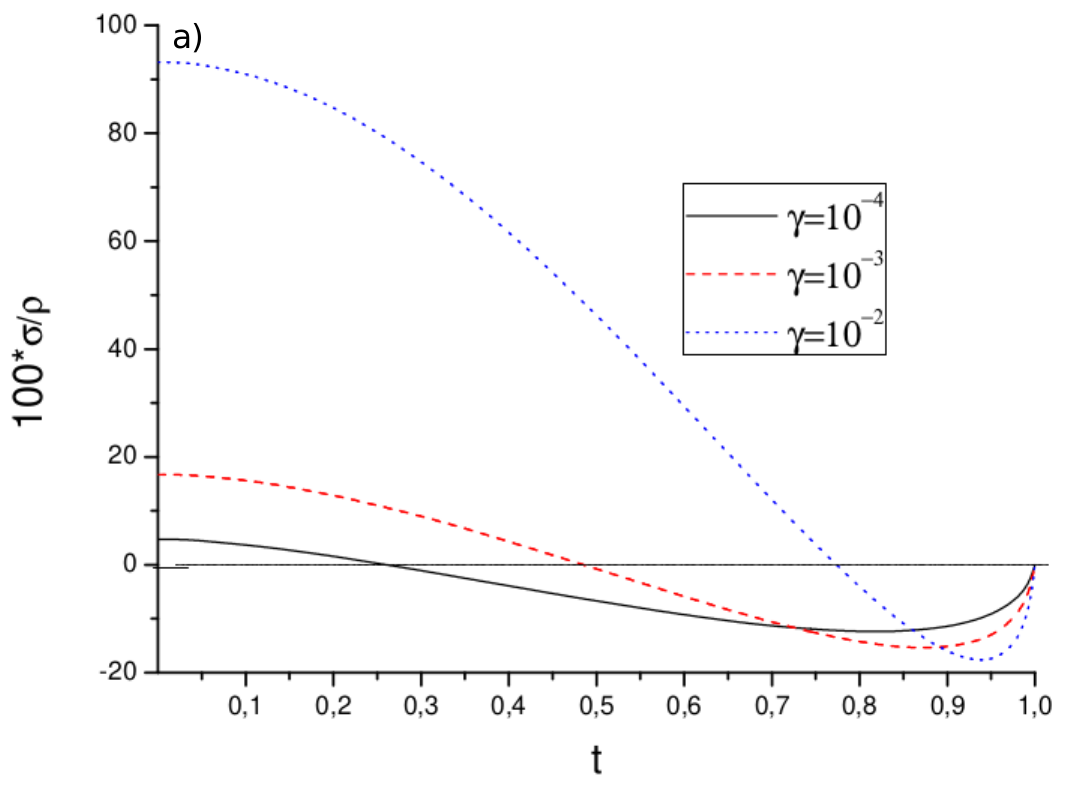}}		
\end{minipage}
\hfill
\begin{minipage}[h]{0.48\linewidth}
\center{\includegraphics[width=0.95\linewidth]{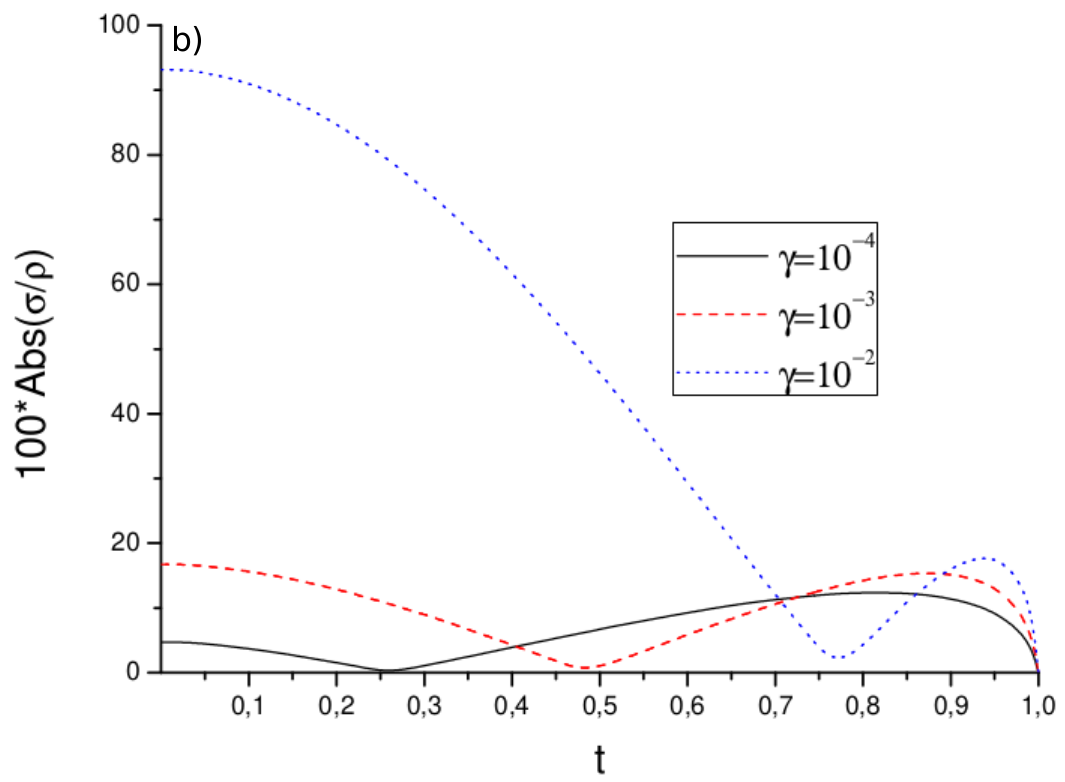}}		
\end{minipage}
\hfill
\caption{$\sigma/\rho$\; (a) and  $\vert\sigma\vert/\rho$\; (b) vs reduced temperature $t=T/T_c$ for three values of the gas parameter:$\gamma=10^{-4},\,\gamma=10^{-3},\,\gamma=10^{-2}$.}
\label{Fig2}
\end{figure}

Below we discuss the origin of non - monotonic behavior of $\absol{\sigma}$. Actually, the equation  \eqref{sigmasx12} for $\xi^2=1$, \,$X_2=0$, \; $X_1=\Delta/2$, has the form:	
\begin{equation}
\sigma=-\frac{\Delta}{V}\sum\limits_k
\left[
\frac{1}{2E_k}+\frac{1}{E_k(e^{E_k \beta} -1)}
\right]\equiv \sigma_0+\sigma_T
\label{sigtt}	
\end{equation}	
where
\begin{equation}
\begin{array}{l}
\sigma_0=-\displaystyle\frac{\Delta}{2V}\displaystyle\sum\limits_k \displaystyle\frac{1}{E_k}=
-\displaystyle\frac{\Delta}{2}\displaystyle\int\displaystyle\frac{d{\bf k}}{(2\pi)^3}\displaystyle\frac{1}{E_k}\\
\\
\sigma_T=-\displaystyle\frac{\Delta}{V}\displaystyle\sum\limits_k\displaystyle\frac{f_B(E_k)}{E_k}
\label{sigtt0}
\end{array}
\end{equation}	
and $E_k$ is given by Eq. \eqref{dispdel}. For a uniform atomic Bose gas the momentum integral in $\sigma_0$ is divergent. This failure can be cured by the dimensional regularization \cite{ourkimyee}, or equivalently by introducing an appropriate counter term: $1/\varepsilon_k$. Then the integral can be taken analytically:
\begin{equation}
\sigma_0=-\frac{\Delta}{4\pi^2}\int\limits_{0}^{\infty}dk k^2 
\left[\frac{1}{E_k}-\frac{1}{\varepsilon_k}\right]=\frac{(\Delta m )^{3/2}}{\pi^2}\ge 0
\label{sigtt0int}
\end{equation}	  
Therefore, $\sigma$  is presented as the   sum of  two terms
with opposite signs:
\begin{equation} 
\sigma=\frac{(\Delta m )^{3/2}}{\pi^2}-\frac{1}{2\pi^2}
\int\limits_{0}^{\infty}\frac{dk k^2}{E_k(e^{E_k \beta} -1)} 
\label{sigtt0inttot}
\end{equation}	  
The first term, due to $\Delta\sim g\rho$, dominates at very low temperatures,  while the second, negative one, prevails at higher temperatures. As a result, at any given input  parameters $(\gamma,t)$ the sign of the function $\sigma(\gamma,t)$ will be determined by the sum of these two competing terms. Particularly, it may cross the abscissa changing the sign, as it is illustrated  in Figs. \ref{Fig2}(a). It is seen  that the intersection point shifts toward high temperatures by  increasing $\gamma$. Such non monotonic  behavior of $\sigma$, leads , naturally, to emergence of extrema in its absolute values $\vert\sigma\vert$ (see Fig. \ref{Fig2}(b)), which can be observed experimentally.     

The question arises, whether such behavior of $\sigma$ will affect the monotony 
of functions $n_0(\gamma,t)$ or $s(\gamma,t)$ vs temperature.  

\begin{figure}[t]
\begin{minipage}[h]{0.32\linewidth}
\center{\includegraphics[width=0.95\linewidth]{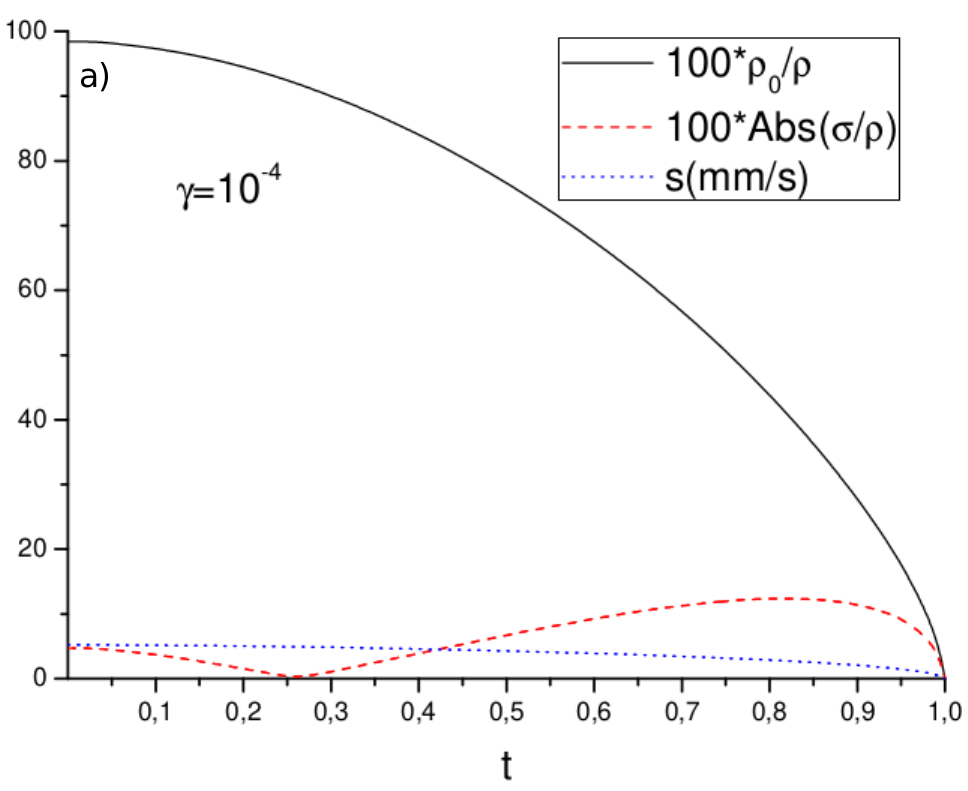}}		
\end{minipage}
\hfill
\begin{minipage}[h]{0.32\linewidth}
\center{\includegraphics[width=0.95\linewidth]{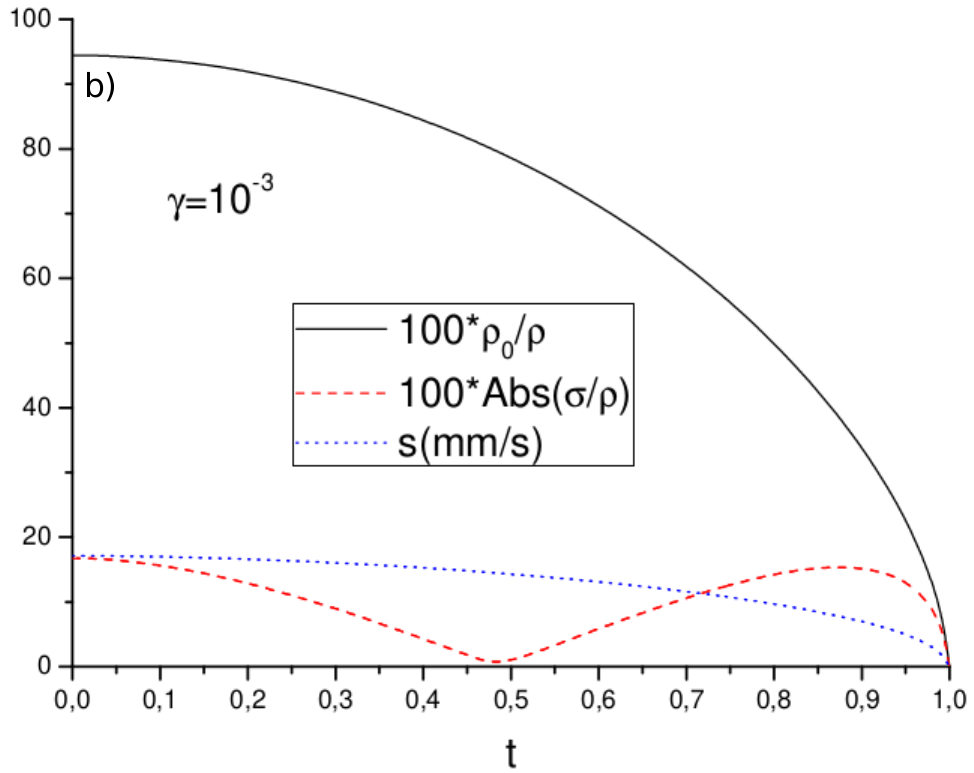}}		
\end{minipage}
\hfill
\begin{minipage}[h]{0.32\linewidth}
\center{\includegraphics[width=0.95\linewidth]{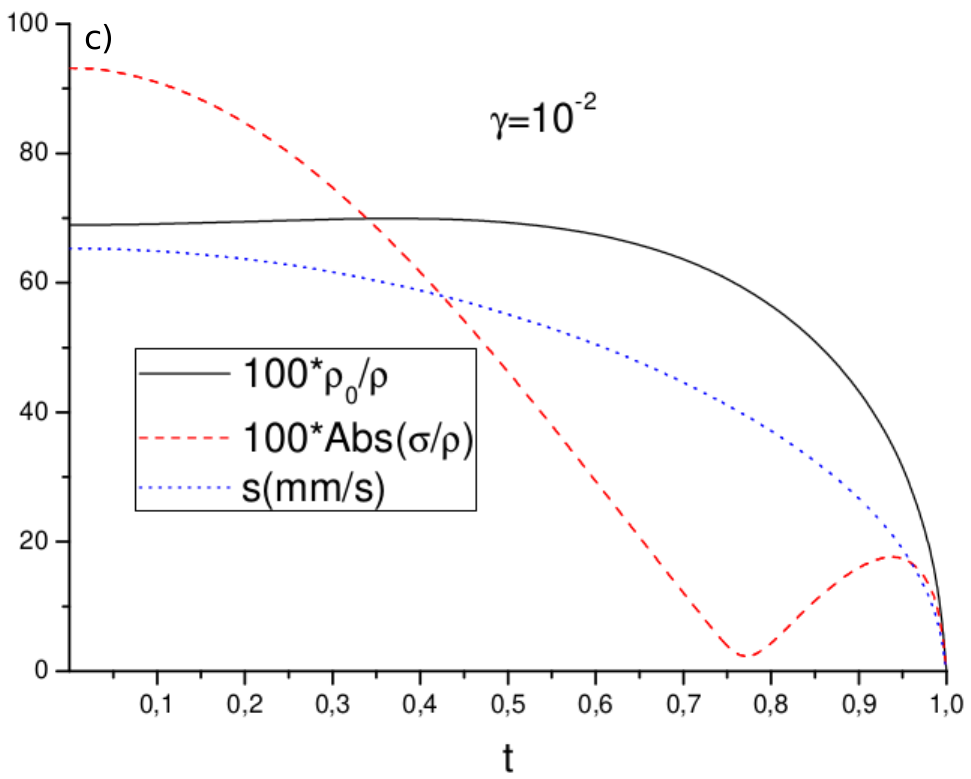}}		
\end{minipage}
\hfill
\caption{The condensed fraction $n_0=\rho_0/\rho$ (solid curve), the fraction of anomalous density $\vert m_1\vert=\vert\sigma\vert/\rho$ (dashed curve) and the sound velocity (dotted curve) vs reduced temperature. Figs. (a), (b) and (c) correspond to $\gamma=10^{-4},\,\gamma=10^{-3}$ and $\gamma=10^{-2}$, respectively.}
\label{Fig3}
\end{figure}

In Figs.\ref{Fig3} we plot the condensed fraction $n_0=\rho_0/\rho$ (solid curve), the fraction of anomalous density $\vert m_1\vert=\vert\sigma\vert/\rho$ (dashed curve) and the sound velocity (dotted curve) vs reduced temperature $t$.  As it is seen from Figs. \ref{Fig3}(a) and \ref{Fig3}(b), in spite of the existence of extrema in   $\vert\sigma\vert$ , the condensed fraction as well as sound velocity remain a  monotonously decreasing function of temperature. This is because, at small values of the gas parameter the condensed fraction  is much larger than $\vert\sigma\vert$, so the latter can be neglected in the main equation \eqref{deleq}, which will be simplified as
\begin{equation} 
\Delta=g\rho_0+g\sigma\approx g\rho_0=g\rho(1-n_1) 
\label{deleqsimpl}
\end{equation}
where $n_1$ is still formally given by the equation \eqref{settteq}. This particular case of  HFB theory is referred in the literature as Hartree - Fock - Popov (HFP) approximation and widely used to study  homogeneous condensates in atomic gases \cite{andersen,salaspopov} and quantum magnets \cite{zapf,nikuni}. On the other hand, as it is shown in Fig. \ref{Fig3}(c) for large $\gamma$ (e.g. $\gamma \ge 10^{-2}$) the anomalous density  becomes rather significant, especially at low temperatures. This means that the application of HFP approximation in this region becomes unjustified. 
 
Coming back to the Figs.\ref{Fig1} it is worth to mention that, experimental data on non - condensate fraction can be also well described in the framework of improved Hartree - Fock - Bogoliubov (IHFB) theory \cite{vanhuy}. However, unfortunately, the authors of this work have not studied the status of anomalous density in their recently proposed theory.

At the end of present section we  raise the issue 'What is the validation of present theory, HFB, with respect of the gas parameter $\gamma=\rho a_s^3\;?$'. Actually, comparison of HFB calculations with Monte -- Carlo ones have revealed that, HFB theory is able to accurately describe, say, $n_0(\gamma, T=0)$ for $\gamma \le 0.1$ \cite{rossi}. Extension of present theory for larger values of $\gamma$, i.e. for stronger interactions, can be performed in the framework of Optimized Perturbation Theory (OPT), developed in works  \cite{yukopt}, \cite{stev42}, \cite{yukdelta1}, since, in this sense, HFB approximation  corresponds to the first order of perturbation in  auxiliary parameter of OPT\footnote{In the literature one may find  equivalent terminology as ``Delta expansion'' or  ``Variational perturbation theory''.} $\delta$. Besides, in this way a possible modification of the critical temperature $\Delta T_c/T_c^{0}=(T_c-T_c^{0})/T_c^{0}$ should be also taken into account. Nevertheless, obviously, our conclusion e.g. on the sign of $\sigma$ remains true. This work is on progress.

\section{Conclusion}
\label{conclusion}	

Actually, present work consists of two parts. In the first part we developed HFB theory  for uniform atomic Bose system for the case when the phase of the condensed wave function is arbitrary. We have shown that, extended Hugenholtz -- Pines relation and the stability condition impose  a certain constraint to such arbitrariness. Namely, the phase angle $\theta$ of a {\underline { homogeneous}} BEC  should satisfy the equation $\sin 2\theta=0$. The sign of the anomalous density, being directly related to the phase is not also observable. However, its absolute value can be estimated uniquely. Physical observables, including $\vert\sigma\vert$ predicted by this conserving versus of HFB theory, clearly, does not depend on the phase $\theta$, as expected.

In the second part we have shown that, the experimental data on the fraction of the condensate in a uniform box can be nicely described within HFB theory. The theory establishes a mathematical relation  between the sound velocity and the anomalous density. However, a physical mechanism by which anomalous density affects the speed of sound remains still not transparent. We calculated also the sound velocity and the absolute value of the anomalous density for finite temperatures. These numerical results serve as a prediction for future experimental studies. 

\section*{Acknowledgments}
We are very grateful to V. Yukalov for his critical comments and also to S. Mamedov for useful  discussions.

\appendix
\section{The main equation in dimensionless form}
\label{app1}
		
Our main equation 
\begin{equation}
\Delta=g(\rho_0+\sigma)
\label{deltaeqapp}
\end{equation}
takes following form in dimensionless variables, $Z=\Delta/g\rho$, $\gamma=\rho a_s^3$ and $t=T/T_c$:
\begin{equation}
Z+n_1(Z)-m_1(Z)-1=0,
\label{zdeltaeqapp}
\end{equation}
where $T_c $ is given by Eq. \eqref{tc}. Here the reduced densities, $n_1=\rho_1/\rho=1-\rho_0/\rho$, $m_1=\sigma/\rho$  are given by
\begin{equation}
\begin{array}{l}
n_1=n_1(T=0)+n_1(T), \quad m_1=m_1(T=0)+m_1(T) 
\label{n1m1app}
\end{array}
\end{equation}
Zero temperature values of densities can be evaluated analytically as:
\begin{equation}
\begin{array}{l}
n_1(T=0)=\displaystyle\frac{1}{2\rho V}\sum_k \left[\displaystyle\frac{\varepsilon_k+\Delta}{E_k}-1\right]= \displaystyle\frac{8Z^{3/2}\sqrt{\gamma}}{3\sqrt{\pi}},\\
m_1(T=0)=-\displaystyle\frac{\Delta}{2\rho V}\sum_k \left[\displaystyle\frac{1}{E_k}-\displaystyle\frac{1}{\varepsilon_k}\right]= \displaystyle\frac{8Z^{3/2}\sqrt{\gamma}}{\sqrt{\pi}}
\label{n1m1appt0}
\end{array}
\end{equation}
To evaluate $n_1(T)$ and $m_1(T)$:
\begin{equation}
\begin{array}{l}
n_1(T)=
\displaystyle\frac{1}{\rho V}\sum_k \displaystyle\frac{\varepsilon_k+\Delta}{E_k}f_B(E_k),\\
m_1(T)=-\displaystyle\frac{\Delta}{\rho V}\sum_k\displaystyle\frac{f_B(E_k)}{E_k}
\label{n1m1appt}
\end{array}
\end{equation}
it is convenient to introduce the dimensionless variable $x=\varepsilon_k/T_c$ in the integral
\begin{equation}
\sum_k  f(k^2)=\frac{4\pi V}{(2\pi)^3}\int\limits_{0}^{\infty} f(k^2) k^2 dk
\label{suminapp}
\end{equation}
As a result $n_1(T)$ and $m_1(T)$ can be presented as
\begin{equation}
\begin{array}{l}
n_1(T)=\displaystyle\int\limits_{0}^{\infty}dx \displaystyle\frac {\sqrt{2x{\tilde c}}f_B(E_x)
}
{2\pi^2E_x\gamma^{2/3}}[x{\tilde c}\gamma^{2/3}+4Z\pi\gamma]\\
\\
m_1(T)=-\displaystyle\int\limits_{0}^{\infty} dx\displaystyle\frac { 2\sqrt{2x{\tilde c}}Z\gamma^{1/3}f_B(E_x)
}
{\pi E_x}
\\
\\
f_B(E_x)=\displaystyle\frac{1}{e^{E_x/t}-1}, \quad
 E_x=\displaystyle\frac{\sqrt{x}\sqrt{x {\tilde c}+8Z\pi\gamma^{1/3}}}{\sqrt{\tilde c}}
\label{n1m1tttapp}
\end{array}
\end{equation}
where  $\tilde c=3.3125$. For the relative  sound velocity, $s/c$ we have
\begin{equation}
\displaystyle\frac{s}{c}=\displaystyle\frac{2(\hbar c)\sqrt{\pi\gamma Z}}
{a_s(mc^2)}
\label{svel}
\end{equation}
where $c\approx 0.3 \times 10^{12}(mm/s)$  is the speed of sound.

\end{document}